\documentclass[prl,twocolumn,floatfix]{revtex4-1}
\usepackage{graphicx}
\usepackage{amsmath}
\usepackage{amsfonts}
\usepackage{subfigure}
\usepackage{txfonts}
\usepackage{xcolor}
\usepackage{comment}

\usepackage{color}
\usepackage[colorlinks,citecolor=blue]{hyperref}

\newcommand{\beq}{\begin{equation}}
\newcommand{\eeq}{\end{equation}}
\newcommand{\bea}{\begin{eqnarray}}
\newcommand{\eea}{\end{eqnarray}}
\newcommand{\bpm}{\begin{pmatrix}}
\newcommand{\epm}{\end{pmatrix}}
\newcommand{\bef}{\begin{figure}}
\newcommand{\eef}{\end{figure}}

\begin{document}
\title{Helical Majorana fermions at the interface of Weyl semimetal and d-wave superconductor: Application to Iridates and high-Tc Cuprates}
\author{Yige Chen$^1$, and Hae-Young Kee$^{1,2,}$}
\email{hykee@physics.utoronto.ca}
\affiliation{Department of Physics, University of Toronto, Ontario M5S 1A7 Canada}
\affiliation{Canadian Institute for Advanced Research, CIFAR Program in Quantum Materials, Toronto, ON M5G 1Z8, Canada}
\date{July 8, 2017}
\begin{abstract}
Majorana bound states exist inside a core of half-quantum vortex in spin-triplet superconductors.  
Despite intense efforts, they have not been discovered in any spin-triplet superconductor candidate materials.
After the success of topological insulator, 
another route to achieve Majorana fermions has been suggested: heterostructures of $s$-wave superconductors and topological insulator or semimetal
with strong spin-orbit coupling provide 
an effective spinless $p+ip$  pairing which supports Majorana bound states in a single vortex. 
This theoretical observation has stimulated both experimental and theoretical communities to search for Majorana fermions, and recently a localized
Majorana state at the end of one-dimensional wire has been reported. 
Here we study the two-dimensional interface of Weyl semimetal and $d$-wave superconductor which promotes a pair of helical Majorana fermions
propagating along the edge of the interface. We suggest that Iridium oxide layer IrO$_2$ classified as two-dimensional Weyl semimetal 
in close proximity to high temperature Cuprates would be a best example to explore these helical Majorana modes. 
\end{abstract}
\maketitle

{\color{blue} \it Introduction.}
Self-conjugate fermions called Majorana fermions (MFs)\cite{Majorana1937} occur
 inside a core of half-quantum vortex in spin-triplet with equal-spin pairing or in a single vortex of
 spinless $p+ip$ pairing states.\cite{Volovik1976, Kopnin1991, Ivanov2001}
It is worthwhile to note that this $p$-wave superconductor (SC)
belongs to the same topological class as the Pfaffian quantum Hall state where the excitations 
are half-quantum vortices with non-Abelian statistics.\cite{Moore1991,Read2000}
These non-Abelian statistics of MFs promote their potential use as a topological quantum qubit\cite{Tewari2007}, 
and the search for MFs has regained some interest.
Recently, discoveries of MFs were claimed: localized Majorana bound states were theoretically proposed at both ends of quantum wire~\cite{LutchynPRL2010, OregPRL2010, FlensbergPRL2015} and further reported
at the edge of topological superconducting wire made of ferromagnetic atom chain on the surface of superconducting lead.\cite{Mourik2012, Rokhinson2012, Lee2014, Yazdani2014, Xu2014}
The experimental set-up was motivated by the theoretical finding that the proximity-induced s-wave superconducting pairing 
on the spin-momentum locked surface states of a topological insulator (TI) is effectively spinless $p+ip$.\cite{FuPRL2008}

The success of TI has not only lead to a discovery of MFs, but
also extended classes of MFs beyond localized bound states.
Analog to the edge states of two-dimensional (2D) quantum spin Hall insulator,   
a pair of counter-propagating MFs protected by time reversal symmetry (TRS) could occur along the edge of
2D systems. These are called helical MFs. 
It was suggested that helical MFs exist at
a one-dimensional (1D)  line junction of  $s$-wave SC, TI, and $s$-wave SC where the 
SC pairing should change its sign across the junction~\cite{FuPRL2008}. 
They have been also proposed to emerge in time-reversal invariant topological SC~\cite{QiPRL2009, HaimPRB2014, TewariPRB2013,SunSciRep2016,NakosaiPRL2012}, 
%time-reversal broken topological SC~\cite{},
%a doped correlated quantum spin Hall insulators
% the bilayer Rashba system~\cite{NakosaiPRL2012}, 
and a 2D Rashba layer proximity to $s_{\pm}$-wave superconductor~\cite{FanPRL2013}. 
These proposals require either bulk time-reversal invariant topological SC, 
or the sign change of $s$-wave pairing potential across the junction or inside the SC.
%which are difficult to achieve in the experimental set-up. 
%The experimental challenge is changing the sign of s-wave pairing potential across the junction. 

\begin{figure}[t]
\centering
\includegraphics[width=9.3cm]{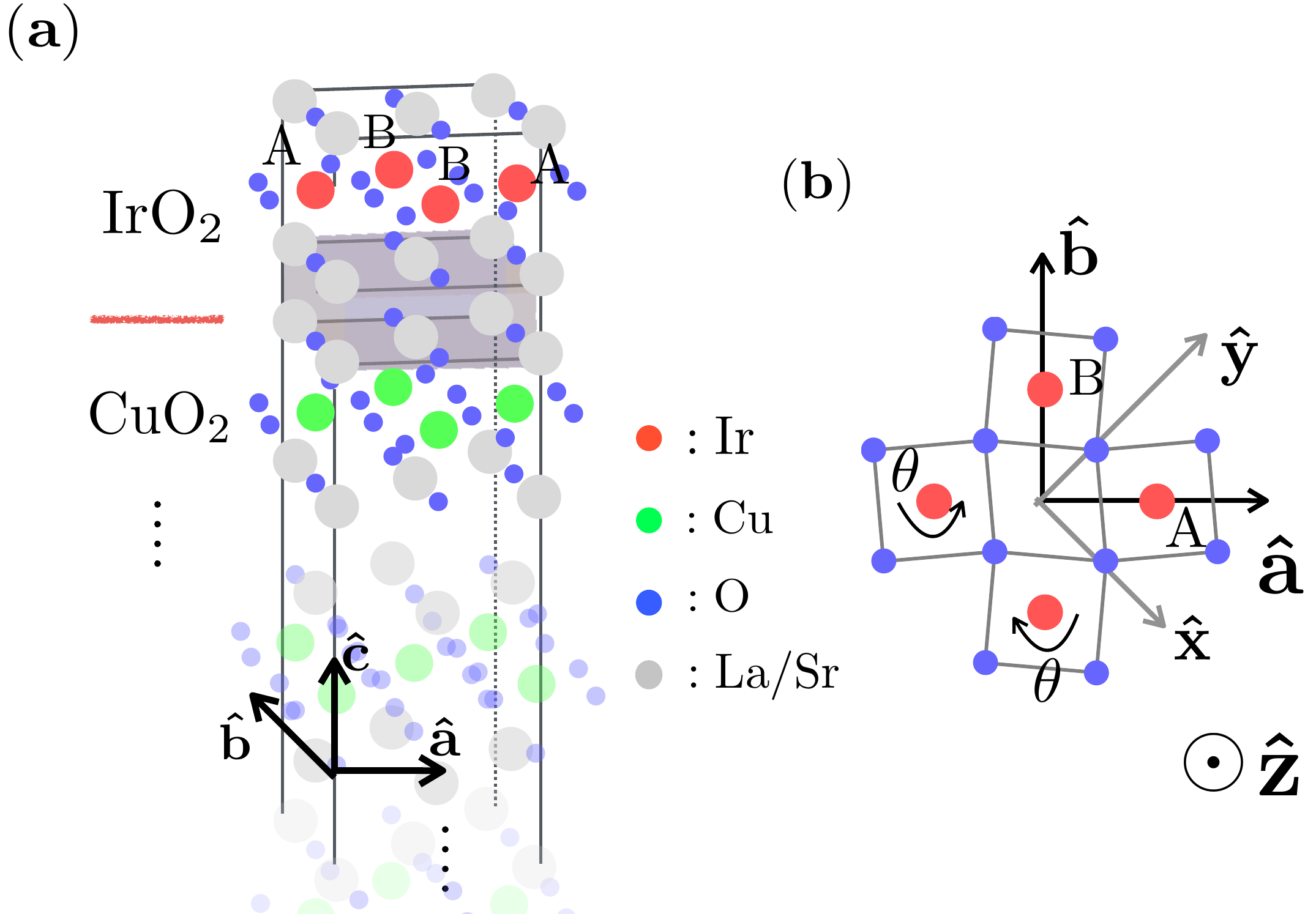}
\caption{(color online) (a) The superlattice structure of IrO$_2$ and high Tc Cuprates (e.g., LSCO shown)  with two Iridium A/B in the unit.
(b) Oxygen octahedra of Iridium have alternating rotations around (001) axis with angle $\theta$. 
$\hat{x}$ and $\hat{y}$ are the Cartesian coordinate defined on the pseudo-square lattice due to the alternating rotation $\theta$. 
The 2D Bravais lattice vectors $\hat{a}=\hat{x}+\hat{y}$ and $\hat{b}=\hat{y}-\hat{x}$.}
\label{fig:1}
\end{figure}

In this paper, we suggest a different mechanism to generate a pair of helical MFs.
We prove that a pair of helical MFs appear at the interface of a 2D Weyl semimetal and d-wave SC.
Here the d-waveness naturally provides
the sign change of the pairing potential across the node and the edge states manifest topological nature of underlying Weyl semimetal.
%Inversion symmetry breaking generated by the interface is essential. 
Since the interface breaks
the inversion symmetry,  a 2D Dirac semimetal in lieu of a 2D Weyl semimetal also supports the same helical MFs.
%where the inversion symmetry perpendicular to the interface is externally broken.
%which is stable for a wide range of parameter space, even though the 2D layer itself has an inversion symmetry. 
Our theory is generally applicable to the interface made of d-wave SCs and Weyl semimetals, but
we focus on the most promising candidate:  IrO$_2$ layer classified as Weyl semimetal with effective pseudospin (J$_{\rm eff}$=1/2)\cite{Carter2012, William2013, YChen14, Rau2016}  
in proximity to d-wave high temperature (high $T_c$) Cuprates such as hole-doped  La$_{2-x}$Sr$_x$CuO$_4$ (LSCO) or YBa$_2$Cu$_3$O$_{7-\delta}$(YBCO).
This also offers a way to induce high temperature SCs in Iridates, which has been a long sought in the community of correlated Mott physics with strong spin-orbit coupling (SOC).\cite{Fa2011, KimNP2016, JKimPRL2012}

{\color{blue} {\it Model.}}
The superlattice structure shown in Fig.~\ref{fig:1}(a) can be experimentally fabricated by growing Sr$_2$IrO$_4$ film on high Tc Cuprates using molecular beam epitaxy or pulsed laser deposition techniques ~\cite{Matsuno2015}. 
The Hamiltonian for IrO$_2$ layer, including the Rashba SOC and proximity effects due to high $T_c$ SC, is given by 
$H=\sum_{\bf k}\psi^{\dagger}_{\bf k} \mathcal{H}({\bf k}) \psi_{\bf k}$: 
\bea
\mathcal{H}({\bf k})&\equiv&\mathcal{H}_{\rm Ir}({\bf k}) \rho_z + \Delta_{\bf k}\rho_x,\; {\rm where}\nonumber\\
\mathcal{H}_{\rm Ir}({\bf k})&=&\epsilon_{\bf k}^0 \tau_x + \epsilon^{\prime}_{\bf k} {\bf I} + \epsilon^{r}_{\bf k} \sigma_z\tau_y +\bold{f}_{\bf k} \cdot \boldsymbol{\sigma} \tau_x+\bold{f}_{\bf k}^{\prime}\cdot \boldsymbol{\sigma}, \nonumber\\ 
\Delta_{\bf k}&=&\Delta({\bf k})\tau_x
.
\label{eq:ham1}
\eea
Here $\psi^{\dagger}_{\bf k}=(c^{\dagger}_{{\bf k},A,\uparrow}, c^{\dagger}_{{\bf k},A,\downarrow},c^{\dagger}_{{\bf k},B,\uparrow},c^{\dagger}_{{\bf k},B,\downarrow},c_{{\bf k},A,\downarrow},-c_{{\bf k},A,\uparrow} ,c_{{\bf k},B,\downarrow},-c_{{\bf k},B,\uparrow})$ 
where $c^{\dagger}/c_{{\bf k}, A/B, \uparrow/\downarrow}$ denotes the electronic creation/annihilation operator to create/destroy an electron with crystal momentum ${\bf k}$ at $A/B$ site and the pseudospin $\uparrow/\downarrow$ in J$_{\rm eff}=1/2$ basis.  
${\boldsymbol \sigma}$, ${\boldsymbol \tau}$ and $\boldsymbol{\rho}$ are Pauli matrices  for J$_{\rm eff}=1/2$,  $A/B$ sublattice, and particle-hole subspace, respectively. 
The 2D crystal momentum ${\bf k}\equiv \left(k_x,k_y\right)$ defined in the pseudo-square lattice.
The Bravais lattice vectors  $\hat{a}=\left(\hat{x}+\hat{y}\right)$ and $\hat{b}=\left(\hat{y}-\hat{x}\right)$ due to the staggered rotations of IrO$_6$ as shown in Fig. ~\ref{fig:1}(b).
The nearest neighbour (NN) intra, inter-orbital, and next-nearest neighbour (NNN) hopping terms with strength $t$, $t^{r}$, and $t^{\prime}$
are given by $\epsilon^0_{\bf k}=2t\left(\cos k_x+\cos k_y\right)$, 
 $\epsilon^{r}_{\bf k}= 2t^{r}\left(\cos k_x +\cos k_y \right)$,
 $\epsilon^{\prime}_{\bf k}=t^{\prime}\cos k_x \cos k_y$, respectively. 
 %in the bulk Hamiltonian of Iridium monolayer $\mathcal{H}_{\rm Ir}({\bf k})$. In particular, $\epsilon^{1d}_{\bf k}$ term appears due to the straggered rotation pattern of local oxygen octahedron structure\cite{YChen14}.  
The NN and NNN Rashba SOC terms occur as ${\bf f}_{\bf k}=2t_R \left(-\sin k_y,\sin k_x,0\right)$ and ${\bf f}^{\prime}_{\bf k}=4 t^{\prime}_R \left(-\cos k_x \sin k_y, \sin k_x\cos k_y,0 \right)$, respectively.
These are  due to the broken inversion symmetry along $\hat{z}$-direction, and can be further enhanced by external electric field. 
%; the side view of superlattice structure is shown in Fig.~\ref{fig:1}(c). 
%Note that the Rashba SOC is due to electric field perpendicular to the 2D interface 
The Cooper pairing potential induced by the proximity effect of high Tc Cuprates manifests $d_{x^2-y^2}$-wave pairing: $\Delta({\bf k})=\Delta_0(\cos k_x-\cos k_y)$ where $\Delta_0$ 
is the strength of pairing. 
%Above is the part is describing the model of Hamiltonian.

\begin{figure}[t]
\centering
\includegraphics[width=8.8cm]{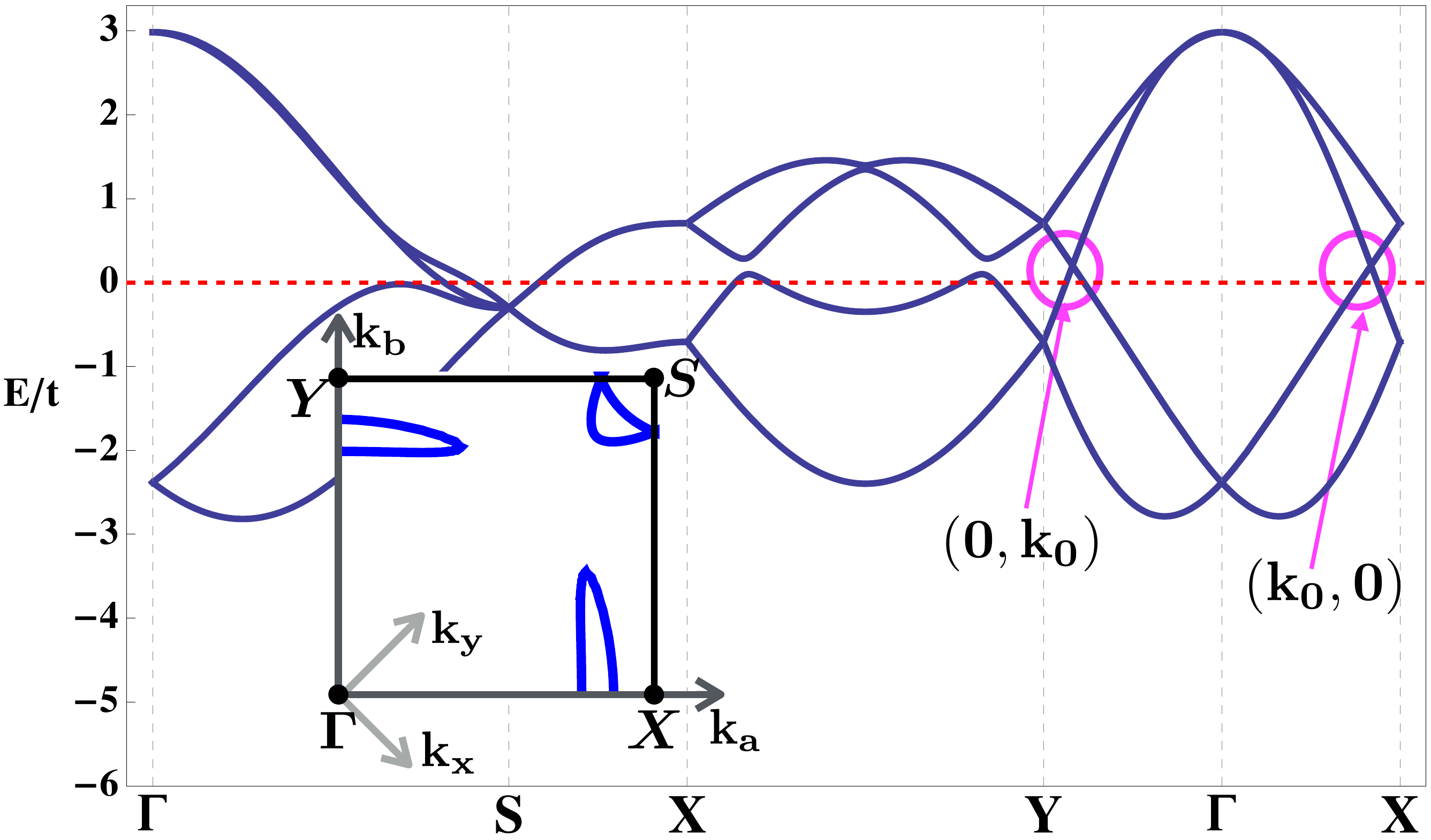}
\caption{(color online) The J$_{\rm eff}$ =1/2 band structure  when $\Delta =0$. The chemical potential $\mu$ is at half filling with hopping parameters $(t^{\prime},t^r,t_R,t_R^{\prime})/t=(-0.5,-0.5,0.4,-0.2)$. There are two pairs of 2D Weyl points, circled pink color, along $\Gamma X$ and $\Gamma Y$. These 2D Weyl points are protected by the glide symmetry and TRS, and the system belongs to symmetry enforced semimetals~\cite{KanePRB2016}. Inset: The FS of the one-quarter of reduced 2D Brillouin Zone (BZ) is shown with reciprocal lattice vectors $k_a=k_x+k_y$ and $k_b=k_y-k_x$.}
\label{fig:2}
\end{figure}

%Symmetry analysis of bulk Hamiltonian before superconducting 
{\color{blue}\it  Symmetry of Iridium oxide layer.}
To determine the existence of symmetry protected edge states,
 let us first investigate the symmetry of IrO$_2$ layer itself. 
When $\Delta_0=0$, $\mathcal{H}_{\rm Ir}({\bf k})$  exhibits symmetry enforced 2D Weyl semimetal~\cite{KanePRB2016}. Unlike their three-dimensional partner,
 these are protected by time-reversal and crystalline symmetries~\cite{KanePRB2016}: 
%The underlying symmetry properties associated with the bulk 2D states are highly nontrivial. 
there are two glides perpendicular to $\hat{a}$ (or $\hat{b}$) directions, which protect the pair of Weyl nodes together with TRS. 
The Rashba SOC is necessary to generate the Weyl nodes. Without such inversion breaking term, the Weyl nodes collapse into time-reversal invariant momentum (TRIM) points, and it becomes a Dirac semimetal. 
The energy dispersion is displayed in Fig.~\ref{fig:2}. Note that, there are two Weyl points at $\left(k_{a/b}=k_0, k_{b/a}=0\right)$ along $\Gamma X$/$\Gamma Y$ high-symmetry line. In the rest of the paper, we use the reciprocal lattice vectors $k_a=k_x+k_y$ and $k_b=k_y-k_x$ to represent the 2D crystal momentum. Under this notation, TRIM points $X/Y\equiv \left(k_{a/b}=\pi,k_{b/a}=0\right)$ and $S\equiv (k_a=\pi,k_b=\pi)$ (see the inset of Fig.~\ref{fig:2}).  

The Bloch bands are doubly degenerate along both $S X$ and $S Y$ protected by a product of TRS
and glide symmetries.  In other words, the product of glide, $\hat{G}_a$ (or $\hat{G}_b$), and time-reversal operator,
 $\Theta_a\equiv \hat{G}_b\hat{T}$ (or $\Theta_b\equiv \hat{G}_b\hat{T}$) becomes the antiunitary operator along the $k_a=\pi$ (or $k_b=\pi$) line, which
 ensure the Kramer degeneracy .  
 Besides, a four-fold rotational axis along $\hat{z}$-direction guarantees that the Fermi surface (FS) is symmetric respect to the interchange of $k_a$ and $k_b$ 
 and thus we show only 1/4 of the FS  in the inset of Fig.~\ref{fig:2}.
When chemical potential $\mu/t=0$ (half filling), the FS made of ${\rm J}_{\rm eff}=1/2$ consists of two hole-pockets around $X$ and $Y$, which are related to the 2D Weyl nodes, and electron-pockets near $S$.~\cite{footnote17}.

%Band structure and Fermi surface (Figure 2)

\begin{figure}[t]
\centering
%\subfigure{
%\includegraphics[width=5.5cm]{Figure3a.pdf}
%}
%\subfigure{
\includegraphics[width=9.5cm]{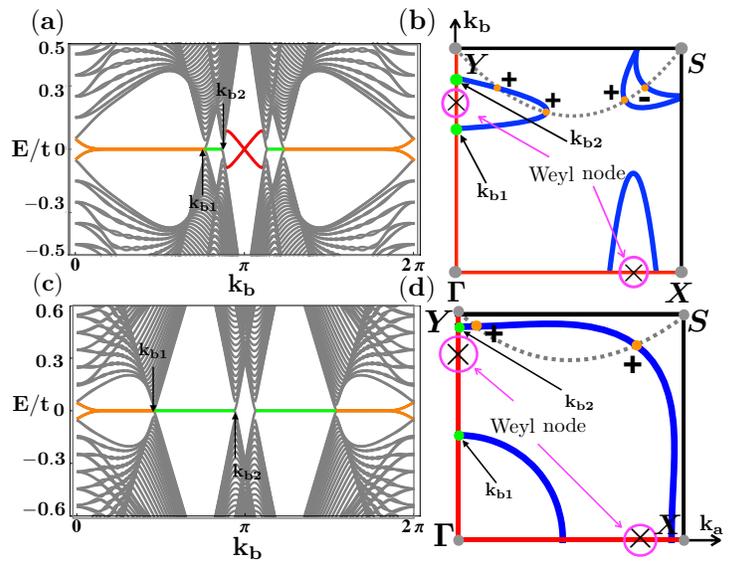}
%}
\caption{(color online) The edge spectrum for (a) $\mu/t=0$ and (c) $\mu/t=1.0$. A pair of {\it helical} gapless modes (red), doubly (orange) and singly (green) degenerate zero-energy flat bands between $0 < k_b < k_{b_1}$ and $k_{b_1}< k_b < k_{b_2}$, respectively, are shown in (a). On the other hand, in (c), only the flat bands exist.
The underlying FS of ${\rm J}_{\rm eff}=1/2$ Iridium colored by blue in the one quarter of first BZ  is shown for (b) $\mu/t=0$ and (d) $\mu/t=1.0$.
Red lines in both (b) and (d) along $k_a$ and $k_b$-axis denote the SC nodes where $\Delta({\bf k})$ vanishes. 
Hence the nodal bulk excitations occur at the intersecting points $k_{b_1}$ and $k_{b_2}$ denoted by green dots in (a) and (c).
The crossed marks refer to the Weyl points near the Fermi energy shown in Fig. 2.
See the main text for $\pm$ signs which determine the topological invariance.}
\label{fig:3}
\end{figure}

%edge state
{\color{blue} {\it One-dimensional Edge states and Majorana Kramer Pairs.}}
Let us investigate the proximity effects of d-wave SC on this 2D symmetry enforced Weyl semimetal of Iridium oxide layer. The bulk band structure of $\mathcal{H}({\bf k})$ remains gapless at two points $(k_{a/b}=k_{b_1},k_{b/a}=0)$ and $(k_{a/b}=k_{b_2},k_{b/a}=0)$ along $\Gamma X$/$\Gamma Y$ direction (highlighted with green color in Fig.~\ref{fig:3}(b) and (d)) because the d-wave node is along $\Gamma X$ and $\Gamma Y$ in the reduced Brillouin zone (BZ).  The symmetry properties associated with the bulk Bloch states hint various exotic excitations on the boundary of the system, which we will discuss in details later.
Below we first show the MFs on the 1D boundary using the slab geometry of superlattice.
%ajorana zero modes which are the essential results in this work. In order to demonstrate the appearance of nontrivial edge excitations, we will first numerically calculate the energy spectrum when the system has open boundary and later identify their topological properties. 

The slab geometry of superlattice has an open-boundary along $\hat{a}$ direction but is periodic along $\hat{b}$-axis, and thus the crystal momentum $k_b$ parallel to the sample edge is a good quantum number. The gapless bulk excitations, mentioned above are projected onto the edge momentum at $k_{b_1}$ and $k_{b_2}$, respectively.
%
%The one-dimensional (1D) edge states exhibit rich exotic gapless excitations at half-filling ($\mu/t=0$). First, The Bloch states (see dispersion with red color in Fig.~\ref{fig:3}(b)) at each boundary, bounded between another projected bulk node at $k_b=k_{b2}$ and $k_b=\pi$, form a pair of {\it helical} gapless modes and support doubly degenerate MZMs, dubbed as Majorana Kramer Pairs (MKPs) at $k_b=\pi$, which is, in fact, related to the TRS~\cite{FanPRL2013}. 
%MKP discussion
%Here the chemical potential $\mu$ is another important parameter. 
When the system is at half filling, there exists a pair of propagating MFs bounded
between $k_{b_2}$ and $- k_{b_2}$: see the linear dispersion centered at $\pi$ with red color 
in Fig.~\ref{fig:3}(a). 
The existence of these helical MFs depends on the relative strength of $\mu$ and $|t_R|$ of NN Rashba SOC term.
% which should be finite due to the broken of inversion symmetry perpendicular to the $xy$-plane. Furthermore, the 2D Weyl nodes, ensured by the TRS and nonsymmorphic symmetry,  
Note that the Weyl nodes at $(\pm k_0,0)$ and $(0, \pm k_0)$ occur only with a finite $t_R$, which simultaneously generates a band gap at $X$ (and $Y$)-points. 
%Recall that a Dirac point should exist at $X$-points without $t_R$. 
When the chemical potential lies inside the band gap at $X$-point, the helical MFs are emerged. 
%
%. $E_{k_0}$  as well, and they play an essential role in inducing such MKPs in the Iridium Oxide layer system. 
Mathematically, when the chemical potential $\mu$ is within the range $-2\sqrt{2} |t_R|  < \mu < 2\sqrt{2} |t_R|$, corresponding to the direct gap of $\mathcal{H}_{\rm Ir}({\bf k})$ at $X/Y$ TRIM points, a pair of helical MFs centered at $k_b=\pi$ appears at the boundary of system. 

%The appearance of MKP is deeply related to the presence of 2D Weyl nodes. 
This should be captured by checking the Berry phase of bulk spectrum and corresponding topological invariance.
% and the matrix element of 
The FS along $\Gamma Y$ of bulk layer forms a closed loop, enclosing the underlying Weyl node, and its Berry phase is $\pi$.
% as displayed in Fig.~\ref{fig:3}(b). Its Berry phase is equal to $\pi$. 
On the other hand,  the two FSs, enclosing $S$ point, have $\pm \pi$-Berry phase.
 % induced by Rashba SOC.  
 In the presence of the time-reversal invariant SC pairing, it was shown that
 % the such Berry phases determine
  the topological invariance is determined by the 
 matrix element of  SC pairing operator between the Kramer pair of Bloch states, i.e., 
 $\delta_{n}(k_{\mathcal{L}})\equiv \langle n, k_{\mathcal{L}}|\hat{T}\Delta^{\dagger}({\bf k})|n, k_{\mathcal{L}}\rangle$.~\cite{Qi2010}
 The time-reversal invariant path $\mathcal{L}$, displayed as the dotted line in Fig.~\ref{fig:3}(b), is picked so that it's invariant under ${\bf k} \rightarrow -{\bf k}$ and moves from $S \rightarrow Y \rightarrow S$. Note that the 1D Hamiltonian on the path $\mathcal{L}$ has both time-reversal (${\hat T}$) and particle-hole (${\hat P}$) symmetries, hence belongs to the topological class $DIII$, which supports a  $\mathbb{Z}_2$ classification in 1D.
Then a 1D $\mathbb{Z}_2$ topological invariant $\nu_{1D}$ is defined as~\cite{Qi2010} 
\beq
\nu_{1D}=\prod_s \left({\rm sgn}\left(\delta_{s}\right)\right),
\label{eq:z2}
\eeq
where the product of $s$ is over all the Fermi points at which  the FS meets the momentum $k_{\mathcal{L}}$ along the contour $\mathcal{L}$.
As shown in Fig. \ref{fig:3}(b), $\delta_{s}$ is positive at two $k_{\mathcal{L}}$s of the FS along $\Gamma Y$ displayed by $+$ sign,
but it has negative sign at one of the two different FS enclosing $S$ point.
This confirms that $\nu_{1D}=-1$, which suggests there exists edge states.
Due to the nature of Weyl metal, a pair of edge states are {\it helical} MFs within the projected regime of path ${\mathcal{L}}$ onto 
$\hat{b}$-direction. A proof of TRS protected MFs at $k_b=\pi$ is presented in Supplemental Materials.

In reality, Iridium layer in close proximity to Curpates is not at half-filling. Because of potential difference, IrO$_2$ will be effectively hole doped.
However, when $|\mu| < 2 \sqrt{2} |t_R|$, the above results hold. On the other hand, when the chemical potential (or doping effect) 
is bigger than the band gap at $X$/$Y$ point generated by the Rashba SOC, a pair of propagating MFs disappears, but the flat zero modes remain as shown in Fig. ~\ref{fig:3}(c).
 % When hole doping the system far away from the half-filled, the system, on the other hand, shows distinct characters in the edge states dispersion. The highlight difference of $\mu/t=1.0$ case from $\mu/t=0$ is the absence of MKPs at $k_b=\pi$ and a pair of 1D helical propagating Majorana modes, as shown in Fig.~\ref{fig:3}(d). Since the chemical potential $\mu/t$ under the current parameter set is over the critical value $2\sqrt{2} t_R/t$,
The disappearance of MFs happens
because the Bogoliubov quasi-particle at $k_b=\pi$ experiences gap closing and re-opening process, accompanying a jump of  $\mathbb{Z}_2$ number. 
It corresponds to the case when the path $\mathcal{L}$ intersects even times with the outermost FS (see Fig.~\ref{fig:3}(d)). Hence the system no longer 
 supports a pair of helical MFs at the boundary, which is consistent with the slab calculation result shown in Fig.~\ref{fig:3}(c). The finding implies that the Iridium layer SC goes into a phase transition to another topological SC by changing $\mu/t_R$, which can be further tuned by an external electric field and/or chemical doping. 

%Topological invariants.

%Slab spectrum with mu=0 and 1.0  

{\color{blue} {\it Zero-energy flat bands}} 
In addition to the helical MFs, the edge spectrum exhibits zero-energy flat states within some regime of $k_b$, localized on the boundary of the system.
The existence of zero-energy flat modes is independent of the chemical potential and the size of Rashba SOC.
 In the regime (a) $0 <k_b < k_{b_1}$, there exist doubly degenerate zero-energy edge states at every $k_b$. 
 On the other hand, the zero-energy edge modes transit from double to single degeneracy in the region (b) $k_{b_1} < k_b < k_{b_2}$. 
%The change of degeneracy associated with edge states is induced when $k_b$ across the projected gapless bulk excitations along $\Gamma X$ at $k_b=k_{b1}$ as shown in Fig.~\ref{fig:3}(b). %The zero-energy flat bands in (a) and (b) regions are due to the chiral symmetry, which maps between the Bloch states with the same energies but opposite sign~\cite{RyuPRL2002}. 

The zero-energy flat bands, emerging within the range  $(0, k_{b_1})$ and $(k_{b_1}, k_{b_2})$ are due to a chiral symmetry,
and characterized by a different topological invariant from $\nu_{1D}$. Note that, the 2D Hamiltonian can be viewed as a collection of 1D Hamiltonian, parametrized by the edge momentum $k_b$, {\it i.e.} $\mathcal{H}({\bf k})\equiv \mathcal{H}_{k_b}(k_a)$.~\cite{RyuPRL2002} The topological class for such 1D Hamiltonian at $k_b \neq \pi$ is $AIII$, and an integer $\mathbb{Z}$ 1D winding number $N_{1D}$ is suitable to identify the bulk-edge correspondence:
\beq
N_{1D}=\frac{1}{2\pi i} \int d k_a {\rm Tr} \left[ \hat{C} \mathcal{H}_{k_b}(k_a)^{-1} \partial_{k_a} \mathcal{H}_{k_b}(k_a) \right],
\label{eq:wn}
\eeq 
with chiral symmetry operator $\hat{C}\equiv \hat{T} \times \hat{P}$ and integrating along $k_a$ momentum. The 1D winding number $N_{1D}=+1$ corresponds the region, highlighted with green color in both Fig.~\ref{fig:3}(a) and (c), where only one zero-energy edge mode exists on each boundary. The regime of edge states with orange color in Fig.~\ref{fig:3}(a) and (c), on the other hand, acquire two zero-energy states, which is consistent with the 1D winding number $N_{1D}=+2$. 
Fig. 4 summarizes the phase diagram by tuning the chemical potential. There exist two distinct topological SC states characterized
by two different topological invariants. When the Iridium film is near half-filling, the system possesses both a pair of helical MFs and the zero-energy flat modes.

%Phase diagram 
\begin{figure}[t]
\centering
\includegraphics[width=8cm]{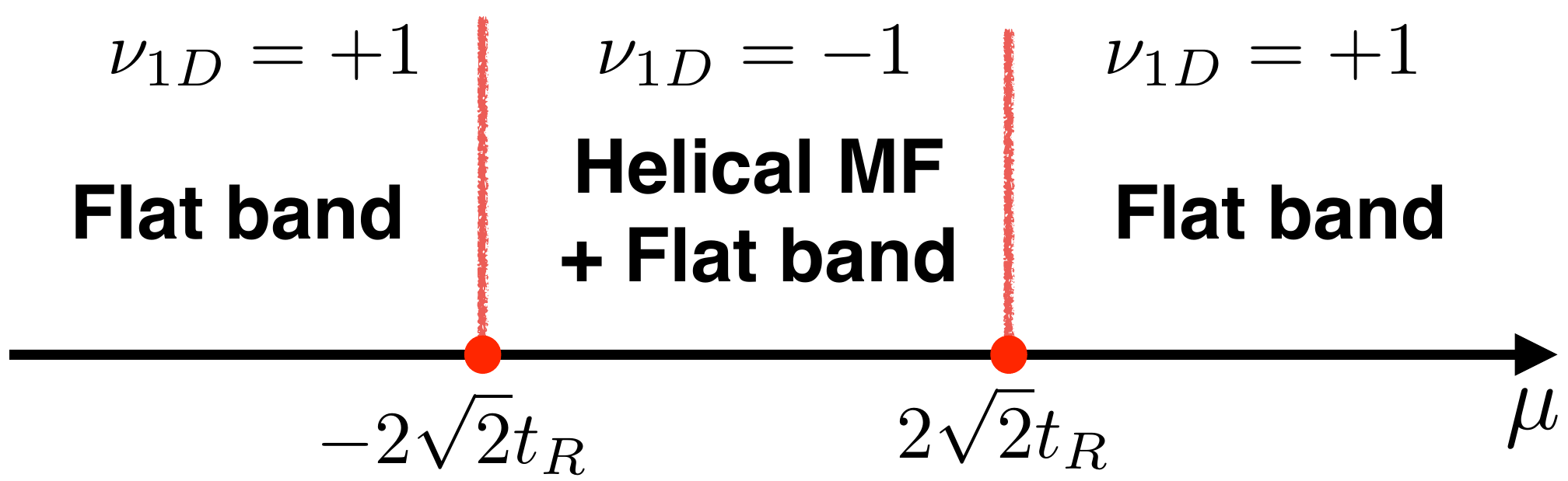}
\caption{(color online) Phase diagram, varying the value of chemical potential $\mu$, indicates that there are two distinct topological SC phases, characterized by the 1D $\mathbb{Z}_2$ number $\nu_{1D}$ and $N_{1D}$. When $\nu_{1D}=-1$, there exists a a pair of helical MFs as well as zero-energy flat bands, while $\nu_{1D}=+1$ case supports only zero-energy flat bands associated with $N_{1D}$.}
\label{fig:4}
\end{figure}
%Discussion about phase diagram 

{\color{blue} \it Discussions.} In this Letter, we have explored the novel Majorana edge spectra of 2D Iridates proximal to $d$-wave SC like
high Tc Cuprates.
%Cooper pairing which is induced through the proximity effect with high-$T_c$ superconductor candidate LCO. 
We find that the 1D edge states exhibit multiple non-trivial features depending on various segment of edge momentum and the amount of doping. Most strikingly, a pair of helical MFs protected by TRS appears along the 1D boundary.
%One can understand the appearance of helical MFs as follows.
Here the d-wave pairing potential vanishes along $\left(k_a,0\right) \& \left(0,k_b\right)$ and changes the sign across these nodal lines.
%It hence effectively generates a domain with opposite sign on $\Delta({\bf k})$.
Since the Cooper pairing on the pseudospin-momentum locked Weyl states
behaves as a $p+ip$ pairing, such a change of pairing potential in d-wave $\Delta({\bf k})$ acts like a line of single vortex, which supports a pair of helical MFs. One should note that the pairing is finite between
the sublattice A and B, thus the MF operator is composed of four fermionic operators as shown in Supplemental Materials. The partners of MF pair are related by TRS to each other. 
%:  $\gamma_1 ({\bf k}_b, E)  = {\hat T} \gamma_2 (-{\bf k}_b, E)$.
%When the octahedra rotation is significant compared to the Rashba SOC, i.e., $t^r > t_R$, these edge MFs have an emergent $S_z$ symmetry.
%In other words,  %$\gamma_{1/2} = \gamma_{\uparrow/\downarrow} = 1/2 (c^\dagger_{\uparrow/\downarrow}+ c_{\uparrow/\downarrow})$, 
%they are analog to MFs of two $A_1$ phases in $^3$He. 
%This resembles a pair of MFs in two copies of $^3$He $A_1$ phase. 
%, similar to the case discussed in Ref.~\cite{FuPRL2008}, which results in the appearance of helical Majorana edge states propagating along the interface when plus non-trivial Berry phase on FS.  

In addition to the helical MFs, the edge spectra in certain momentum regimes 
%of $k_b$ {\it i.e.} $0<k_{b}<k_{b2}$, on the other hand, 
exhibit flat zero energy bands, which resembles  the situation presented in 1D edge states of noncentrosymmetric d-wave superconductor.
These flat edge states require the clean edge shape.
%~\cite{footnote2}. 
For example, the current results were obtained by the translation invariant boundary along $\hat{b}$-axis.
%It appears only along $\hat{a}$- or $\hat{b}$-axis,
%because 
These edge modes are emerged due to the chiral symmetry, and thus they could acquire a slight dispersion when small but finite
 further neighbour hopping paths are taken into account.
On the other hand, the helical MFs 
%are robust against a small perturbation as long as TRS is preserved.
are not sensitive to either the boundary shape nor further neighbour hopping, because they are protected by the TRS.  
However, they disappear when the open boundary is along the $x$ or $y$-direction.
This is because 
 the two Weyl states near $X$ and $Y$ points are projected into the same boundary and
the sign change of d-wave pairing occurs twice for this open boundary, which cancel out
the nontrivial topological effects. 
 %
%S-wave effect
These results are robust in the presence of a small isotropic (or anisotropic)
$s$-wave pairing potential as long as TRS is preserved.
In this case, the weak $s$-wave pairing  potential slightly shifts the location of nodal lines, and thus 
%where the superconducting potential $\Delta({\bf k})$ vanishes. 
 it does not alter the existence of MFs. 
 
 There has been a long sought for a high $T_c$ SC in Iridates.
Since high Tc Cuprates provide both hole doping and proximity to $d$-wave SC, this interface also offers a way to generate a high $T_c$ SC in Iridates, where the amount of doping can be controlled by the thickness of insulating barrier to reduce the potential difference, but preserving a proximity effect.
%
%
%Effective doping deserves some discussion. 
%
In summary,  the helical MFs along the 1D boundary of Iridium oxide layer proximal to high Tc Cuprates exist due to a combination of the d-wave pairing 
and Weyl states at any open boundary except $x$- or $y$-direction. 
Our proposal also offers a way to induce high Tc SC in Iridates in addition to uncovering the helical MFs in correlated oxide systems. 

{\color{blue} \it Acknowledgement.}
%HYK thanks to H. Takagi for useful discussions. 
This work was supported by the NSERC of Canada and the center for Quantum Materials at the University of Toronto.

%%Reference


\begin{thebibliography}{30}
\bibitem{footnote17} Depending on the other higher hopping terms amd contribution from J$_{\rm eff}$=3/2 bands, additional FS near $\Gamma$ can be generated, but their existence does not affect the main conclusion of the current work.
\end{thebibliography}
\end{document}